\documentclass[12pt]{article}
\usepackage{times}
\usepackage{graphicx}
\usepackage{geometry}
\usepackage[utf8]{inputenc}
\usepackage{enumitem,amssymb}
\usepackage{ragged2e}
\newlist{thematic}{itemize}{8}
\setlist[thematic]{label=$\square$}
\usepackage{pifont}
\newcommand{\cmark}{\ding{51}}%
\newcommand{\done}{\rlap{$\square$}{\raisebox{2pt}{\large\hspace{1pt}\cmark}}%
\hspace{-2.5pt}}

\usepackage{jheppub}   
\usepackage[dvipsnames,svgnames,x11names]{xcolor}
\usepackage{sectsty}
\usepackage[outercaption]{sidecap}
\sidecaptionvpos{figure}{c}

\definecolor{DarkGreen}{rgb}{0.0, 0.3, 0.0}
\definecolor{purple}{rgb}{0.5, 0.0, 0.5}
\definecolor{red}{rgb}{1, 0.0, 0.0}
\definecolor{green}{rgb}{0, 1.0, 0.0}

\def\3he{$^3{\rm He}$}

\def\lsim{\mathrel{\lower2.5pt\vbox{\lineskip=0pt\baselineskip=0pt
           \hbox{$<$}\hbox{$\sim$}}}}

\def\gsim{\mathrel{\lower2.5pt\vbox{\lineskip=0pt\baselineskip=0pt
           \hbox{$>$}\hbox{$\sim$}}}}

\begin{document}
\raggedright
\huge
Astro2020 Science White Paper \smallskip

Exploration and characterization of the earliest epoch of galaxy formation: beyond the re-ionization era


\bigskip

\normalsize

\noindent \textbf{Thematic Areas:} \hspace*{60pt} $\square$ Planetary Systems \hspace*{10pt} $\square$ Star and Planet Formation \hspace*{20pt}\linebreak
$\square$ Formation and Evolution of Compact Objects \hspace*{31pt} $\square$
 Cosmology and Fundamental Physics \linebreak
  $\square$  Stars and Stellar Evolution \hspace*{1pt} $\square$ Resolved Stellar Populations and their Environments \hspace*{40pt} \linebreak
$\done$
    Galaxy Evolution   \hspace*{45pt} $\square$             Multi-Messenger Astronomy and Astrophysics \hspace*{65pt} \linebreak
  
\textbf{Principal Author:}

Name:\,Kotaro Kohno
 \linebreak						
Institution:\,Institute of Astronomy, Graduate School of Science, The University of Tokyo, Japan
 \linebreak
Email:\,kkohno@ioa.s.u-tokyo.ac.jp
 \linebreak
Phone:\,+81-422-34-5029
 \linebreak
 
\textbf{Co-authors:~} Yoichi Tamura (Nagoya), 
                    Akio Inoue (Osaka Sangyo U.), 
                    Ryohei Kawabe (NAOJ), 
                    Tai Oshima (NAOJ), 
                    Bunyo Hatsukade (U.~Tokyo), 
                    Tatsuya Takekoshi (U.~Tokyo), 
                    Yuki Yoshimura (U.~Tokyo), 
                    Hideki Umehata (RIKEN), 
                    Helmut Dannerbauer (IAC, ULL), 
                    Claudia Cicone (INAF), 
                    Frank Bertoldi (AIfA)\linebreak

\textbf{Abstract:~} State-of-the-art rest-frame UV and FIR photometric and spectroscopic observations are now pushing the redshift frontiers of galaxy formation studies up to $z\sim9-11$ and beyond. Recent HST observations unveiled the presence of a star-forming galaxy exhibiting the Lyman break at $\lambda_{\rm obs}=1.47\pm0.01$ $\mu$m, i.e., a $z=11.09^{+0.08}_{-0.12}$ galaxy with a stellar mass of $\sim10^9 M_\odot$, demonstrating that galaxy build-up was well underway early in the epoch of reionization (EoR) at $z>10$. Targeted spectroscopy of a lensed Lyman break galaxy uncovers the earliest metals known to date up to $z=9.1096\pm0.0006$ by detecting the bright [OIII] 88~$\mu$m nebular line, indicating the onset of star formation 250 million years after the Big Bang, i.e., corresponding to a redshift of $z\sim15$. These latest findings lead us to a number of key questions: How and when metal enrichment happened in the EoR? What was the nature of the earliest-epoch star-forming galaxies at $z=10-15$? What was the spatial distribution of such galaxies, and what was the relation to the putative large-scale ionization bubbles during the EoR? What were the dark-halo masses of such earliest-epoch star-forming galaxies? To address all these questions, we need to uncover a statistically large number of $z=10-15$ galaxies in the pre-reionization era. 
Here we argue two possible pathways: (1) a wide-area, sensitive blind spectroscopic survey of [OIII] 88 $\mu$m line-emitting galaxies at submillimeter wavelengths, and (2) an ultra-wide-area, high-cadence photometric survey of transient sources at radio-to-(sub)millimeter wavelengths, together with the immediate follow-up spectroscopy with an ultra-wide-band spectrograph, to catch the pop-III $\gamma$-ray bursts. 

\pagebreak

\justify

\section{Introduction: unveiling the cosmic star-formation history}

Unveiling the dust-obscured part of the cosmic star-formation history is one of the biggest challenges in the modern astrophysics. Intensive surveys using airborne and ground-based facilities at (sub)millimeter to far-infrared wavelengths have revealed that the majority of the cosmic star formation at $z = 1 – 3$, where the cosmic star-formation rate density (SFRD) peaks, seems to be obscured by dust (e.g., Burgarella et al.~2013 \cite{Burgarella2013}). However, SFRDs beyond redshift $> 3 – 4$ have been mostly investigated by the rest-frame ultra-violet emission (e.g., Bouwens et al.~2015 \cite{Bouwens2015}; Ono et al.~2018 \cite{Ono2018}), and the roles of the dust-obscured star-formation largely remain unsettled (e.g., Bouwens et al.~2016 \cite{Bouwens2016}; Rowan-Robinson et al.~2016 \cite{Rowan-Robinson2016}). Now ALMA has been routinely used to conduct deep continuum surveys (e.g., Dunlop et al. 2017 \cite{Dunlop2017}; Franco et al.~2018 \cite{Franco2018}; Hatsukade et al.~2018 \cite{Hatsukade2018}) as well as blind spectral scans to uncover CO and [CII] 158 $\mu$m line emitting galaxies (e.g., Aravena et al.~2016 \cite{Aravena2016}; Yamaguchi et al.~2017 \cite{Yamaguchi2017}; Gonz{\'a}lez-L{\'o}pez et al.~2017 \cite{Gonzalez2017}). Furthermore, a large number of targeted ALMA continuum observations have been used to probe the evolution of star-forming interstellar medium up to $z\sim6$ (e.g., Scoville et al.~2016 \cite{Scoville2016}). 

\begin{figure*}[!hb]
\begin{centering} 
\includegraphics[width=0.8\linewidth]{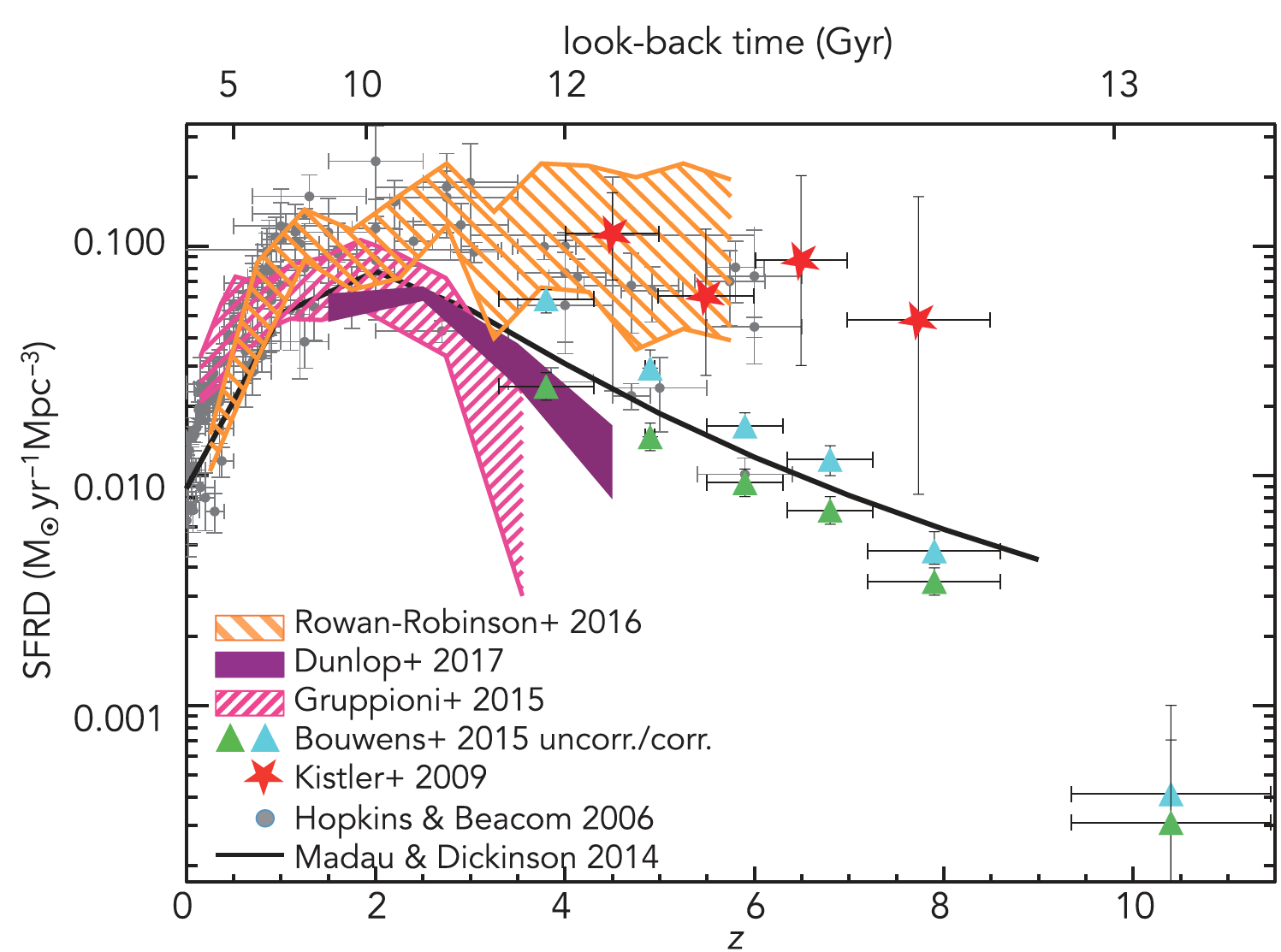}
\caption{The redshift evolution of the cosmic star-formation rate density (SFRD). Different derivations of the obscured and unobscured SFRD are compared. The SFRDs measured based on the rest-frame UV (Bouwens et al.~2015 \cite{Bouwens2015}) show a sharp decline from $z\sim3-4$ to $z\sim10$, whereas the FIR(Herschel)-based SFRDs (Rowan-Robinson et al.~2016 \cite{Rowan-Robinson2016}) and long-$\gamma$-ray-burst (GRB)-based estimation (Kistler et al.~2009 \cite{Kistler2009}) exhibit more flattened SFRDs even up to $z\sim6-8$. Figure taken from Gruppioni et al. (2017) \cite{Gruppioni2017}.
}\label{fig:SFRD}
\end{centering} 
\end{figure*} 


State-of-the-art rest-frame UV and FIR photometric and spectroscopic observations are now pushing the redshift frontiers of galaxy formation studies up to $z\sim9-11$ and beyond. Recent HST observations unveiled the presence of a star-forming galaxy exhibiting the Lyman break at $\lambda_{\rm obs}=1.47\pm0.01$ $\mu$m, i.e., a $z=11.09^{+0.08}_{-0.12}$ galaxy with a detectable amount of stellar continuum emission with a stellar mass of $\sim10^9 M_\odot$, demonstrating that galaxy build-up was well underway early in the epoch of reionization (EoR) at $z>10$ (Oesch et al.~2016 \cite{Oesch2016}). More recently, ALMA observations of the lensed Lyman break galaxy MACSJ1149-JD1 uncovered the earliest metals known to date up to $z=9.1096\pm0.0006$ by exploiting the bright [OIII] 88$\mu$m nebular line (Hashimoto et al.~2018 \cite{Hashimoto2018}). The spectral energy distribution (SED) analysis, combined with the precisely determined redshift, indicates that the dominant stellar component of this galaxy was formed about 250 million years after the Big Bang, i.e., corresponding to a redshift of $\sim15$. Thermal dust emission has also been detected toward galaxies at $z=7.5-8.3$ (Watson et al.~2015 \cite{Watson2015}; Laporte et al.~2017 \cite{Laporte2017}; Tamura et al.~2018 \cite{Tamura2018}), imposing stringent constraints on the dust formation processes in the early universe (e.g., Aoyama et al.~2017 \cite{Aoyama2017}). {\it All these recent state-of-the-art HST and ALMA observations point to the importance of uncovering galaxies beyond $z>10-15$, i.e., during the pre-reionization era.}

\section{Key questions, and what we need next}

These recent discoveries of $z=9-11$ galaxies, demonstrating that the onset of earliest star formation just a few 100 million years after the Big Bang, i.e., $z\sim15$, lead to a number of key questions: 
How and when metal enrichment happened in the EoR? What was the nature of the earliest-epoch star-forming galaxies at $z=10-15$? What was the spatial distribution of such galaxies, and what was the relation to the putative large-scale ionization bubbles during the EoR? What were the dark-halo masses of such earliest-epoch star-forming galaxies?  
To address all these key questions, we need to establish the methodology to uncover a statistically large number of spectroscopically-identified galaxies at $z=10-15$. 

Here we present two potential pathways to go, i.e., (1) a wide-area ($>1$ deg$^2$), sensitive ($\sim$1 mJy, 5$\sigma$) blind spectroscopic survey of [OIII] 88 $\mu$m line-emitting galaxies at submillimeter wavelengths covering $\sim190-360$ GHz, and (2) a ultra-wide-area ($>$a few 100 deg$^2$), high-cadence ($>$once per day) photometric survey of transient sources at radio-to-(sub)millimeter wavelengths, together with immediate follow-up spectroscopy with a high-spectral-resolution ($R>30,000$), ultra-wide-band spectrograph covering e.g., $80-180$ GHz, in order to catch the signals from the pop-III GRBs.

\subsection{A wide-area spectroscopic survey of [OIII] 88~$\mu$m line-emitting galaxies}

The hydrogen Ly $\alpha$ line has been most commonly observed to identify the spectroscopic redshift of high-redshift galaxies. However, the Ly $\alpha$ line becomes increasingly weak at $z>8$ because the neutral fraction of the intergalactic medium increases. Inoue et al. (2014) \cite{Inoue2014} have investigated the use of the rest-frame far-infrared nebular emission lines for determining spectroscopic redshifts of $z>8$ galaxies. They find that the [OIII] 88 $\mu$m line
can be a promising tool for spectroscopy of $z>8$ galaxies based on a cosmological hydrodynamic simulation of galaxy formation (Shimizu et al.~2014), where they incorporated a model of [OIII] emission as a function of metallicity calibrated by FIR observations of local galaxies.
The [OIII]~88 $\mu$m line traces massive star formation, since the ionization of O$^+$ $\rightarrow$ O$^{++}$ requires hard ($E>$35.1 eV) ionizing photons from hot, short-lived O-type stars. Since the first detection of [OIII]~88 $\mu$m toward a bright Lyman $\alpha$ emitter at $z=7.212$ (Inoue et al. 2016) \cite{Inoue2016}, its usefulness has become more evident as demonstrated by e.g., Hashimoto et al. (2018). 

Moriwaki et al. (2018) \cite{Moriwaki2018} used a high-resolution cosmological simulation of galaxy formation and reported the statistics and physical properties of a population of [OIII]~88 $\mu$m emitting galaxies at $z>7$. Figure 2 (left) shows the projected distribution of [OIII]~88 $\mu$m line emitters based on their numerical studies, demonstrating the necessity of a $\sim$1 deg scale survey to construct a statistically large number of $z>7$ star forming galaxies. The evolution of [OIII] 88 $\mu$m line luminosity functions has also been predicted based on the latest UV luminosity functions (Ishigaki et al.~2018 \cite{Ishigaki2018}; Oesch et al.~2018 \cite{Oesch2018}) with fixed faint-end slope and $M^*$ and evolving number density. The luminosity functions shown in Figure 2 (right) suggest that a wide-area ($>1$ deg$^2$), sensitive ($\sim1$ mJy or better, 5$\sigma$) spectroscopy survey is necessary to obtain a statistically large number of [OIII] 88 $\mu$m emitters at $z\sim8$ and to uncover a significant number of candidate $z=10-15$ [OIII] emitters. 

\begin{figure*}[!hb]
\begin{centering} 
\includegraphics[width=0.9\linewidth]{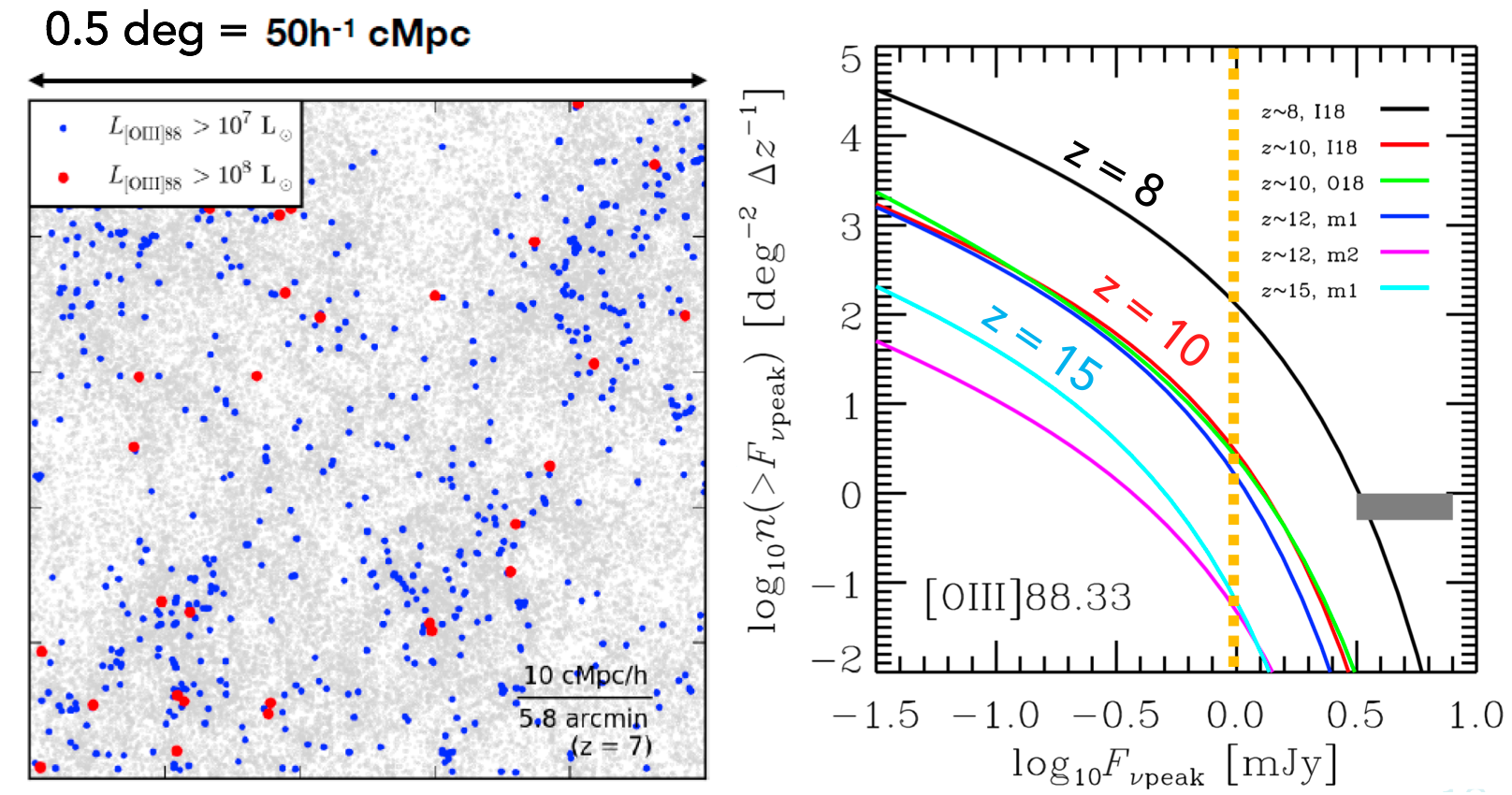}
\caption{(Left) The projected distribution of [OIII] 88 $\mu$m line emitting galaxies at $z=7$ in a cubic volume of comoving 50 $h^{-1}$ Mpc ($\sim$30 arcmin) on a side. The grey points are all galaxies in this volume, whereas the blue and red points represent galaxies with $L_{\rm [OIII] 88} >10^7 L_\odot$ and $>10^8 L_\odot$, respectively. Note that the [OIII] line luminosity of MACSJ1149-JD1 at $z = 9.110$ is $\sim7 \times 10^7 L_\odot$ (Hashimoto et al.~2018). Figure taken from Moriwaki et al.~2018. \cite{Moriwaki2018}. (Right) Predicted [OIII] 88 $\mu$m luminosity functions at $z = $8, 10, 12, and 15. Vertical yellow dashed line indicate the required line sensitivity of 1 mJy. Figure taken from Inoue, A., et al. (private communication).
}\label{fig:OIII}
\end{centering} 
\end{figure*} 

\subsection{An ultra-wide-area, high-cadence photometric survey of the reverse-shock emission from high-redshift long-duration GRBs}

Another unique approach to uncover star-forming galaxies during the EoR is to detect the Synchrotron emission from the reverse shock of the long-duration GRBs. Inoue et al. (2007) \cite{Inoue2007} showed that the reverse shock component of the Pop-III GRBs at $z = 15 - 30$ seen in the $\sim$300 GHz band continuum can be substantially brighter than 1 mJy. In order to catch the reverse shock emission, immediate follow-up observations at a time scale of a few hours to $\sim 10$ hours (i.e., less than 1 day) are necessary. Due to the difficulty of such time-critical observations of GRBs using the existing facilities (including ALMA), the study of the reverse shock emission remains largely unexplored yet. Nevertheless, a recent successful detection of the reverse shock component from GRB 120326A at $z = 1.798$, which was detected by 1.3-mm observations using SMA at $\sim$17 hours after the burst, has demonstrated that such prompt mm-wave continuum measurements can catch the brightest part of the SED from a GRB at a cosmological distance (Urata et al.~2014) \cite{Urata2014}.

\begin{figure*}[!hb]
\begin{centering} 
\includegraphics[width=0.9\linewidth]{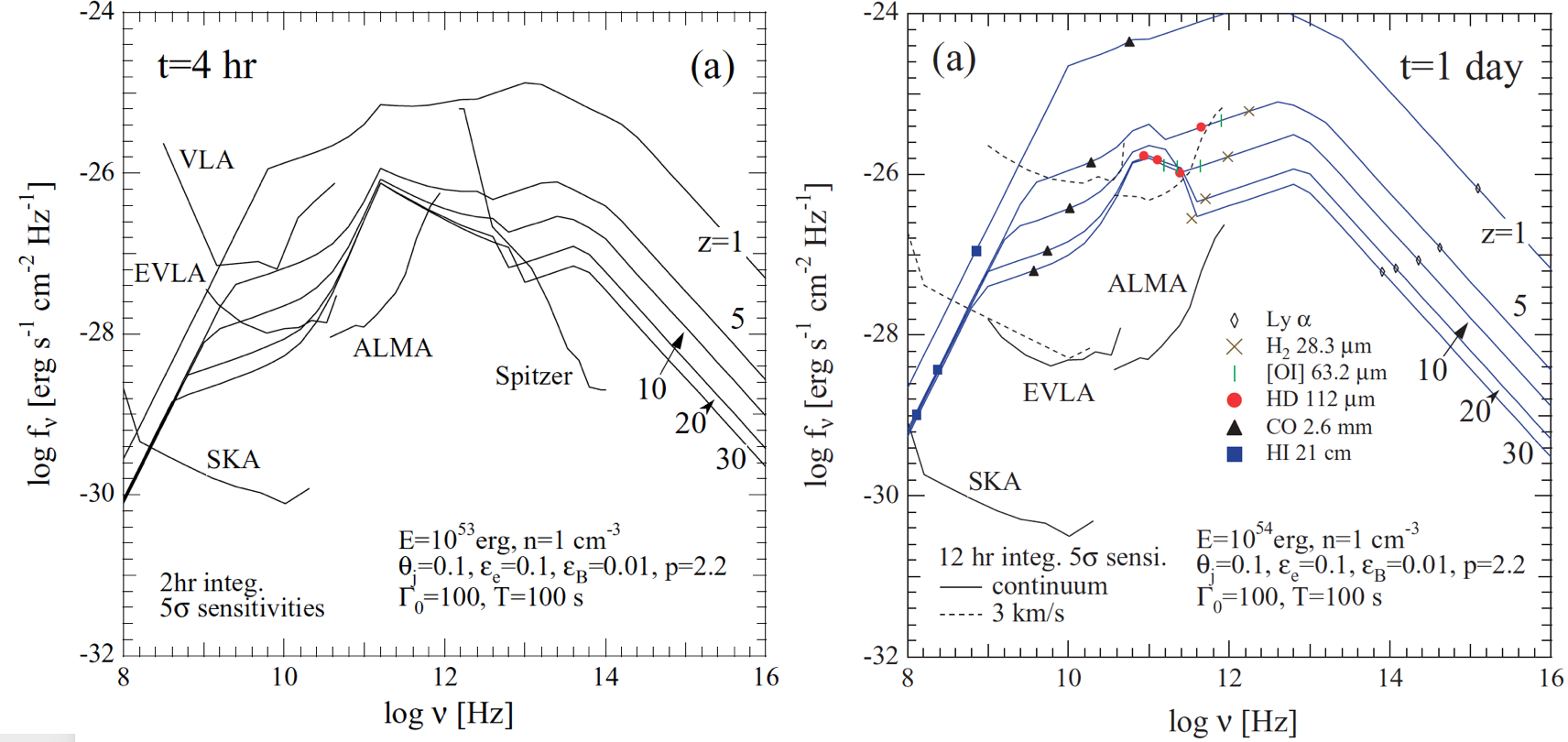}
\caption{(Left) Broad-band spectra of GRB afterglow at different $z$ labelled at fixed post-burst observing time $t$ = 4 hours for a fiducial GRB. Overlayed are 5$\sigma$ continuum sensitivities of various observing facilities, assuming 2 hours integration (i.e., 50\% of the post-burst observer time). (Right) Broad-band spectra of GRB afterglows with fiducial parameters except for $E=10^{54}$ erg, at different $z$ as labelled and fixed post-burst observer time $t$ = 1 day. The redshifted frequencies for the lowest lying transitions of H$_2$, HD, CO, and [OI], as well as the Ly $\alpha$ and HI 21-cm lines are indicated. Note that the next generation radio facilities like ngVLA should also be sensitive enough to search for such reverse shock emission and subsequent spectroscopy follow-up. Figures taken from Inoue et al. (2007) \cite{Inoue2007}.
}\label{fig:GRB}
\end{centering} 
\end{figure*} 

\section{Prospects for the Next Decade}

ALMA will continue to play a central role to extend the spectroscopic redshift frontier by improving its capabilities (i.e., increasing instantaneous bandwidth, sensitivity, and FoVs). Once we can find good targets for spectroscopy, ALMA will be sensitive enough to detect and image the redshifted [OIII] 88~$\mu$m line up to $z\sim20$. However, submillimeter telescopes optimized for wide-area spectroscopy surveys, which allow us to conduct a $>1$-deg$^2$-area survey with a spectroscopic sensitivity of $\sim1$ mJy (5$\sigma$) or better at the submillimeter wavelengths (i.e., $\lambda_{\rm obs}$ = 0.8--1~mm) using a significant amount of time investment ($>$1,000 hours), will be crucial to drastically increase the number of spectroscopically-identified $z>10$ galaxies, given the limited ALMA capabilities for blind surveys. NIRSpec on {\it JWST} will also be sensitive enough to detect star-forming galaxies at $z>10$ by exploiting the bright rest-frame UV line, CIII]1909, although it also offers a modest capability for blind surveys of such CIII]1909 line-emitting galaxies due to its limited FoV. Air-borne survey-optimized facilities at the near-infrared wavelengths covering $\lambda_{\rm obs}$ = 2--5 $\mu$m, which enable a $>100$ deg$^2$ survey with a modest depth ($\sim$27 $m_{\rm AB}$), will be crucial to uncover the promising candidate $z>10$ galaxies for spectroscopic follow-up observations. Note that the existing and planned survey missions, including {\it WFIRST} and {\it Euclid}, will not have a survey capability above $\lambda>2$ $\mu$m.  

\medskip

GRBs are expected to be routinely uncovered while high-energy survey telescopes like {\it SWIFT} are in operation, but they tend to have a large positional errors (up to a few arcmin) especially at the early stage of their detections. These sources may also be faint or invisible in the optical and near-infrared bands. Even without prior alert from $\gamma$-ray telescopes and subsequent multi-wavelengths follow-up, we will be able to blindly detect such (sub)millimeter-wave transient sources once a large-aperture ($D\sim50$ m) single-dish telescope is coupled with a large-format direct detector array which fills $>0.5-1$ deg$^2$ FoV of the telescope: the expected mapping speed can reach $\sim10^3-10^4$ deg$^2$ mJy$^{-2}$ hr$^{-1}$, which is 4--5 orders of magnitude faster than the current ALMA, and still $2-3$ orders of magnitude faster than the Toltech on the LMT 50-m telescope. An ultra-wide-band spectrometer at (sub)millimeter-waves will then be crucial to identify the spectroscopic redshift of the GRB via atomic-line absorption against the bright (sub)millimeter-wave continuum emission produced from the GRB reverse shock. One of the candidates will be the HD line, a primordial molecule. The 2-mm atmospheric window allows us to search for the HD 112 $\mu$m absorption lines for $z = 14-20$, and the 3-mm window corresponds to $z = 22-30$ and beyond. Key advancements required will be (1) instantaneous data analysis of multiple-visit $>100$ deg$^2$ mm/submm observations, (2) sophisticated technique to detect mm/submm transient sources from these images, and (3) immediate fine-resolution ($R>30,000$) spectroscopy follow-up observations with a ultra-wide-band instantaneous bandwidth covering e.g., $\sim 80-180$ GHz. 

\medskip

These ambitious surveys are expected to be implemented by the planned large-aperture ($D\sim50$ m), survey-oriented submillimeter-wave facilities such as the Large Submillimeter Telescope (LST, Kawabe et al.\ 2016 \cite{Kawabe17}), and the Atacama Large-Aperture sub-mm/mm Telescope (AtLAST)\footnote{\url{http://atlast-telescope.org}}, which are supposed to merge into a unified, multi-national project. The proposed spectroscopy surveys of [OIII] line-emitters will also automatically detect a large number ($>10^6$) of CO and [CII] 158 $\mu$m emitters at $z<8-10$, which will have enormous synergies with the next-generation surveys in the optical/near-infrared wavelengths (See Astro2020 white paper by Geach et al.). And more importantly, these spectroscopic surveys will be unique even in the SKA-low era, because it will be impossible to resolve individual galaxies at the EoR in the HI emission. Comparison of the spatial distributions of the individually-detected [OIII] 88-$\mu$m emitting galaxies and the large-scale distributions of HI (SKA-low) and Ly$\alpha$ (e.g., {\it SPHEREx}) will tell us about the reionization history of the early universe. High cadence radio-to-(sub)millimeter imaging surveys will be complementary to those in the optical wavelengths by using e.g., HSC/Subaru and LSST, and we expect to uncover new classes of (sub)mm transient sources such as orphan GRBs (Totani \& Panaitescu 2002) \cite{Totani2002} and wondering massive blackholes (Kawaguchi et al.~2014) \cite{Kawaguchi2014}. 

\pagebreak

\bibliographystyle{unsrturltrunc6}
\bibliography{references}

\end{document}